\documentclass[cameraready]{Interspeech}

\usepackage{cite}

\title{What Makes Synthetic Speech Sound Sarcastic?
A Prosody-Controlled Perception Study} 

\author[orcid=0000-0002-1409-2482]{Zhu}{Li}
\author[orcid=0000-0002-4277-4851]{Shekhar}{Nayak}
\author[orcid=0000-0002-7631-5063]{Matt}{Coler}

\address{
    Speech Technology Lab, University of Groningen, The Netherlands
}

\email{\{zhu.li, s.nayak, m.coler\}@rug.nl}


\usepackage{comment}

\makeatletter
\def\bstctlcite#1{\@bsphack
  \@for\@citeb:=#1\do{%
    \edef\@citeb{\expandafter\@firstofone\@citeb}%
    \if@filesw\immediate\write\@auxout{\string\citation{\@citeb}}\fi}%
  \@esphack}
\makeatother

\begin{document}

\bstctlcite{IEEEexample:BSTcontrol}

\maketitle

\begin{abstract}
Prosody plays an important role in sarcasm perception, yet previous studies have relied on naturally produced speech that lacks fine-grained control over individual acoustic dimensions. As prosodic cues co-vary in natural data, isolating their independent contributions remains challenging. We introduce a controlled framework using neural text-to-speech (TTS) with prompt-based prosodic conditioning to manipulate speech rate, pitch variation, and loudness. An orthogonal stimulus set was constructed to enable causal testing of prosodic cue effects. Human listeners rated sarcasm and naturalness, and their judgments were compared with predictions from a foundation model capable of processing audio input. Results show that loudness primarily drives human sarcasm perception, whereas the model assigns greater weight to speech rate, indicating limited behavioral alignment. This study shows how controllable neural TTS enables investigation of prosodic cue weighting in speech perception. 
\end{abstract}

\section{Introduction}
Sarcasm is a form of verbal irony in which speakers convey an intended meaning that contrasts with the literal content of an utterance. 
Behavioral and neurocognitive evidence suggests that sarcasm comprehension integrates context, utterance content, and affective prosody, with context-content incongruity serving as a primary cue and prosody modulating interpretation in a context-dependent manner \cite{woodland2011context, matsui_role_2016, nakamura_context-prosody_2022}.
At the prosodic level, early work showed that listeners can use naturally produced prosodic cues to recognize verbal irony, especially when utterances are textually ambiguous \cite{bryant2002recognizing}. 
Context-minimal paradigms therefore provide a controlled way to test whether vocal cues alone can bias listeners toward sarcasm and to identify which prosodic features serve as potential cues.
However, such findings do not imply the existence of a single, universal ``sarcastic tone of voice'' \cite{bryant2005there}. 
Instead, when interpreting sarcasm, listeners may attend to multiple auditory dimensions, including pitch modulation, speech rate, loudness, and intonation \cite{voyer2010subjective, voyer_context_2016}. 
Listeners interpret these cues together with linguistic and contextual information \cite{bryant2010prosodic}, and their use is not uniform across listeners, varying with individual differences and contextual incongruity \cite{riviere2018using}.
These perceptual findings are complemented by acoustic analyses showing that sarcastic speech is associated with heterogeneous prosodic patterns, including changes in pitch contours, temporal structure, and vocal intensity \cite{rockwell2000lower, cheang_sound_2008, chen2018s, li2024functional}. 
Cross-linguistic work in Mandarin and Cantonese further suggests that prosodic cues to sarcasm extend beyond English \cite{li_role_2020, lan_acoustic_2025, cheang2011recognizing}. 

Despite these advances, a methodological challenge remains. Many studies rely on naturally produced speech, typically contrasting sarcastic and literal renditions of the same utterance, which provides ecological validity but limits experimental control \cite{rockwell2000lower}. 
Other work has examined spontaneous ironic speech in discourse, showing that irony may be marked by relative prosodic contrast with preceding speech \cite{bryant2010prosodic}. 
However, these approaches do not isolate the causal role of individual acoustic dimensions, because pitch range, speaking rate, amplitude, and other prosodic features often co-vary in natural speech. As a result, post hoc acoustic analyses can describe group-level differences between sarcastic and literal productions, but cannot determine which specific prosodic features independently bias perception. Controlled acoustic manipulation offers a way to address this limitation by varying selected vocal cues while holding other properties constant.

Recent advances in large language model (LLM)-based neural TTS have substantially improved the naturalness and controllability of synthetic speech \cite{hu2026qwen3}. Building on prior work using acoustically manipulated speech to create sarcastic-sounding stimuli \cite{peters2017creating} and recent synthesis-based modeling of sarcasm cues \cite{li2025modeling}, modern speech synthesis systems provide a useful testbed for generating natural-sounding sarcastic speech with explicit control over prosodic attributes such as pitch, timing, and intensity, making synthetic speech increasingly viable as tightly controlled stimuli for psycholinguistic experiments.
This study builds on the context-minimal setting and leverages speech synthesis advances by generating prosody-controlled sarcastic speech with Qwen3-TTS \cite{hu2026qwen3}, a multimodal foundation model that supports controllable speech generation. We manipulate three theoretically motivated prosodic dimensions, speech rate, pitch range, and amplitude, while holding lexical content constant. 
This design allows us to test the individual and combined contributions of prosodic cues to sarcasm perception. Additionally, we complement the human perception experiment with a computational evaluation, assessing how a large-scale model responds to the same stimulus set.
Comparing human and model responses allows us to examine whether computational systems are sensitive to the same acoustic manipulations as human listeners and to identify potential divergences in cue weighting between humans and models.

By integrating controlled synthetic speech, human perception experiments, and machine-based perception, this study proposes a novel framework for investigating speech perception. 
Our results show that human listeners rely primarily on loudness cues when interpreting sarcasm, with louder realizations receiving significantly higher sarcasm ratings than softer ones. 
Model-based evaluation suggests a different pattern. While the foundation model exhibited sensitivity to prosodic variation, its predictions were primarily driven by speech rate, assigning higher sarcasm scores to slowed speech. 
These findings clarify the perceptual role of individual prosodic cues and demonstrate the value of neural TTS for constructing tightly controlled speech stimuli and conducting speech perception experiments.


\section{Method}

\subsection{Materials}

The linguistic materials were adapted from the stimulus set provided in Bryant and Fox Tree \cite{bryant2002recognizing}, which consists of short English utterances originally used to investigate verbal irony perception in spontaneous speech. In their experiments, these sentences were shown to be semantically neutral in isolation and can plausibly convey either a sincere or sarcastic interpretation depending on prosodic realization or context. They served as input for neural speech synthesis to systematically manipulate prosodic dimensions. 

\subsection{Speech synthesis and stimulus selection} 

All stimuli were generated using the publicly released Qwen3-TTS-12Hz-1.7B-CustomVoice model\footnote{ \url{https://huggingface.co/Qwen/Qwen3-TTS-12Hz-1.7B-CustomVoice}} with a single synthetic speaker voice to eliminate inter-speaker variability and uncontrolled voice quality differences. We systematically manipulated three prosodic dimensions in a fully crossed $2 \times 2 \times 2$ factorial design: pitch variation (dynamic vs.\ flat), loudness (loud vs.\ soft), and speech rate (fast vs.\ slow). Prosodic manipulations were implemented via natural-language prompting to the TTS model with explicit instructions controlling tempo, pitch variation, and intensity. For example, in the \textit{slow flat soft} condition, the model was instructed to ``Speak in a very slow, dragging tempo. Keep your voice quiet, almost like mumbling to yourself. Crucially, use a completely flat, monotone pitch with absolutely no emotion or intonation changes.''

For each utterance and condition, 100 candidate samples were generated using stochastic decoding with different random seeds. To modulate prosodic variability, sampling temperature was adjusted across conditions. Temperature controls the sharpness of the output distribution during autoregressive decoding \cite{holtzman2020curious}.
Lower temperature reduced variability in flat conditions, whereas higher values encouraged expressive variation in dynamic conditions.
This procedure yielded a large candidate pool for each utterance and condition, from which stimuli were selected based on the orthogonality criteria described below.

Acoustic features were extracted from all candidates: pitch variation was measured as the standard deviation of fundamental frequency (dynamic: 39.67 $\pm$ 17.03 Hz; flat: 34.96 $\pm$ 14.82 Hz), loudness as mean intensity (loud: 15.71 $\pm$ 9.05 dB; soft: 13.12 $\pm$ 10.62 dB), and speech rate as utterance duration (fast: 2.24 $\pm$ 4.79 s; slow: 3.55 $\pm$ 13.68 s).
%
\begin{figure}[h!]
    \centering
    \includegraphics[width=\linewidth]{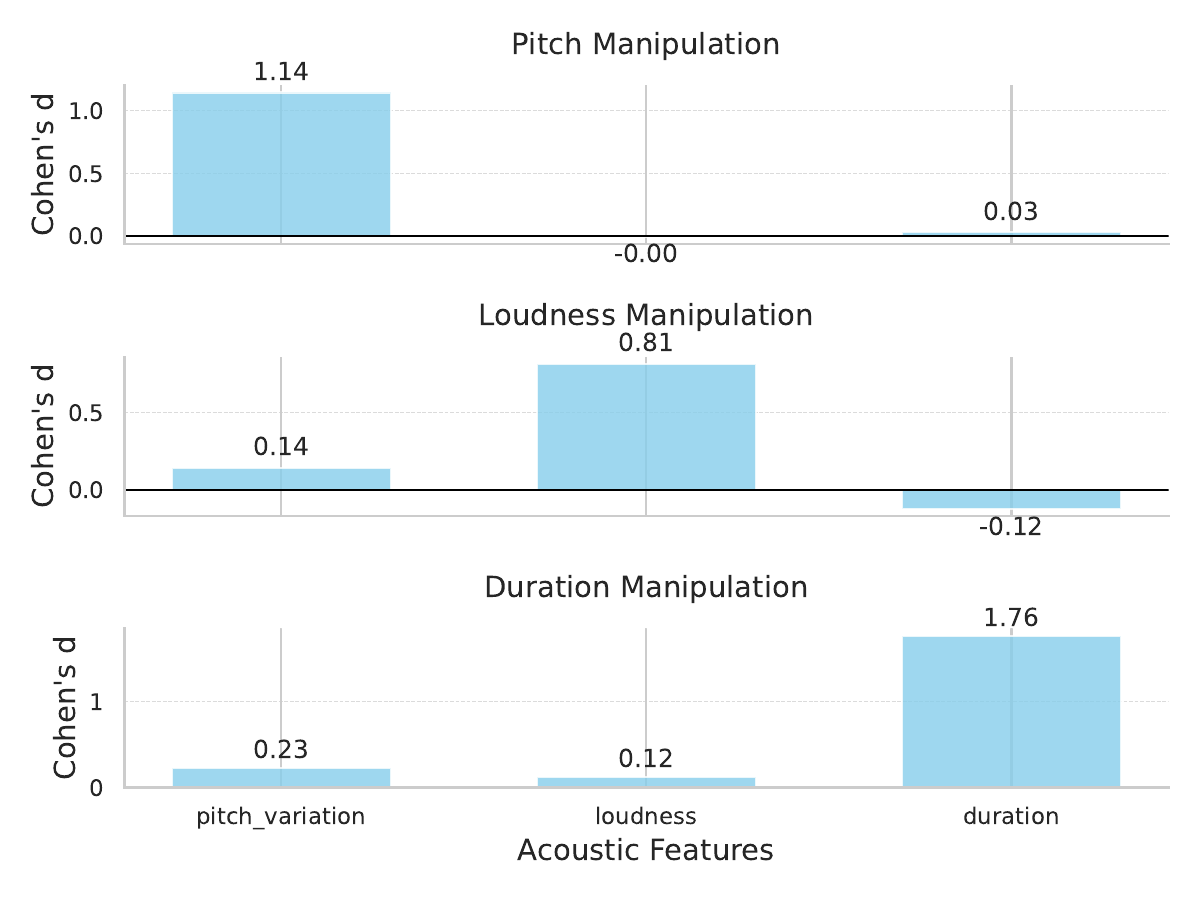}
    \caption{
    Acoustic validation of orthogonal prosodic manipulations.
    Each panel shows mean pairwise effect sizes (Cohen’s $d$) for contrasts along one target dimension.
    Target dimensions exhibit large effect sizes, whereas non-target dimensions show minimal effects,
    confirming independent and orthogonal manipulation of the three prosodic features.
    }
    \label{fig:acoustic_validation}
\end{figure}
To ensure orthogonality across prosodic dimensions, we implemented an effect-size-based stimulus selection procedure. For each utterance and condition pair, candidate samples were contrasted along the relevant factorial dimensions, and pairwise differences were quantified using Cohen’s $d$ \cite{cohen2013statistical} to estimate manipulation strength and potential cross-dimensional spillover onto non-target dimensions. Stimuli were selected to maximize effect sizes in the intended dimension while minimizing effect sizes in non-target dimensions. The candidate combination yielding the optimal orthogonality profile was retained as the representative stimulus for each utterance and condition.
Importantly, since prosodic manipulations were implemented through natural-language prompting of a generative TTS system, complete isolation of single acoustic dimensions cannot be guaranteed. Our orthogonality analysis therefore establishes statistical independence among measured prosodic features, but does not preclude subtle variation in unmeasured dimensions. 
To address this concern, we examined key voice-quality measures, namely the amplitude difference between the first and second harmonics (H1-H2) and the harmonics-to-noise ratio (HNR) across conditions and observed no systematic differences (all $p > 0.05$).
%

This procedure resulted in a final stimulus set exhibiting large effect sizes in the intended dimensions (pitch variation: $d = 1.14$; loudness: $d = 0.81$; duration: $d = 1.76$), while effect sizes in non-target dimensions remained close to zero (all $|d| < 0.25$). These results confirm effective and independent manipulation of the three prosodic features, ensuring a controlled, orthogonal set of prosodic stimuli suitable for the following perceptual experiments. 

\begin{figure*}[h]
\includegraphics[width=\linewidth]{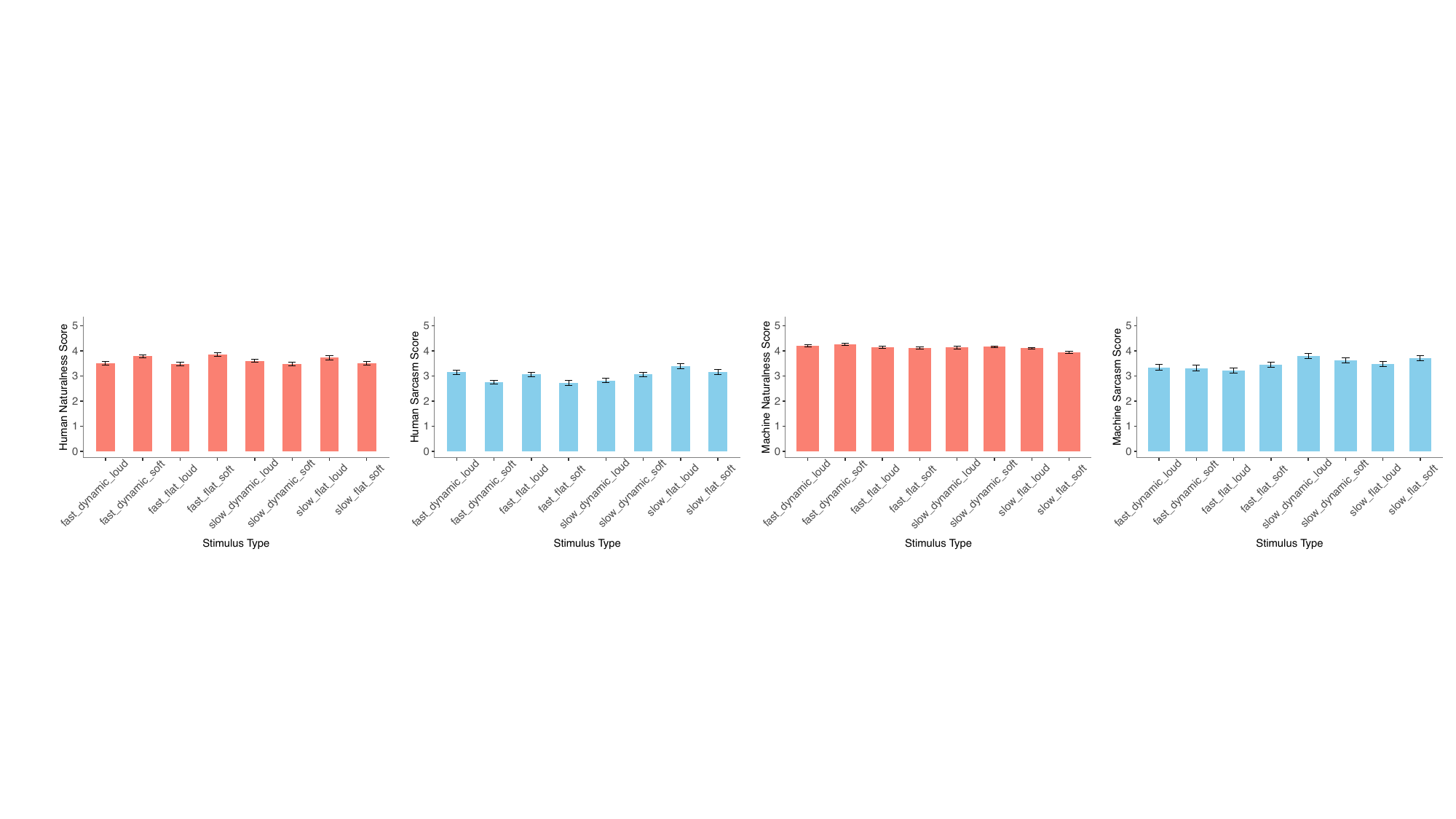}
\centering
\caption{
Mean sarcasm (left) and naturalness (right) ratings across prosodic conditions for humans and machines.
Human ratings are averaged across participants and utterances, whereas machine ratings are averaged across model seeds and utterances.
Error bars represent standard errors of the mean.
}
\label{fig:human_machine_score}
\end{figure*}

\subsection{Experimental design}
The final stimulus set followed a fully crossed $2 \times 2 \times 2$ prosodic design (pitch variation $\times$ loudness $\times$ duration). Because lexical content was held constant across prosodic conditions, any perceptual differences can be attributed primarily to prosodic manipulation; the orthogonality of these manipulations is validated acoustically in Figure~\ref{fig:acoustic_validation}. 
The 192 stimuli (24 utterances $\times$ 8 prosodic conditions) were distributed across eight counterbalanced lists. Each participant was assigned to one list and heard three utterances per prosodic condition (3 $\times$ 8), yielding 24 unique sentences per participant. Within a given list, lexical items did not repeat across prosodic conditions. This design ensured balanced exposure across participants while minimizing fatigue and potential lexical confounds.

For human sarcasm perception, we recruited 66 speakers 
with native or reported near-native proficiency in English who completed an anonymized online questionnaire. Participants listened to the target utterances and rated perceived sarcasm and naturalness on separate five-point Likert scales (1 = not sarcastic / very artificial, 5 = very sarcastic / very natural). No contextual information was provided; judgments were based exclusively on vocal prosody in the synthesized speech.

For model-based sarcasm perception, we used Qwen3-Omni \cite{Qwen3-Omni}, a foundation model capable of processing audio input. The model was instructed to simulate a participant in a speech perception experiment and to base its judgments on vocal prosody. 
For each stimulus, the audio waveform was provided together with a fixed prompt directing the model to evaluate prosodic cues (pitch variation, speaking rate, and loudness). 
The model produced a five-point sarcasm rating, a five-point naturalness rating, a categorical attitude label (“Sarcastic” or “Sincere”), and a brief explanation grounded in prosodic features. To reduce stochastic variability, inference was repeated with six different random seeds, and ratings were averaged across runs.
Stimuli used in the human experiment and presented to the model are identical, enabling a direct comparison between human listeners and the model's behavior.

\subsection{Evaluation and statistical analyses}

All statistical analyses were conducted in \texttt{R}. Linear mixed-effects models were fitted using the \texttt{lmer} function from the \texttt{lme4} package \cite{bates2015fitting}. Speech rate (fast vs.\ slow), pitch contour (dynamic vs.\ flat), and loudness (soft vs.\ loud) were treated as fixed effects, with random intercepts for participants and items. Fast speech rate, dynamic pitch variation, and soft loudness were specified as the reference levels for the categorical predictors. Models were estimated using maximum likelihood.
Post-hoc pairwise comparisons were conducted using estimated marginal means obtained from the \texttt{emmeans} package \cite{lenth2019package}. Multiple comparisons were adjusted using the Tukey method.



\section{Results}
\subsection{Reliability of ratings}
We first evaluated reliability using intraclass correlation coefficients (ICC) \cite{shrout1979intraclass} to justify the use of aggregated human and model ratings in subsequent analyses.
Inter-rater reliability for human sarcasm ratings was modest at the individual level (ICC$_{(2,1)} = 0.15$) but high after aggregation (ICC$_{(2,k)} = 0.92$), indicating that group-averaged judgments provide stable estimates. 
For human naturalness ratings, reliability followed a similar pattern, with low individual agreement (ICC$_{(2,1)} = 0.04$) and improved reliability after averaging (ICC$_{(2,k)} = 0.76$).
For model predictions, reliability across seeds was moderate for sarcasm (ICC$_{(2,1)} = 0.40$; ICC$_{(2,k)} = 0.80$) and lower for naturalness (ICC$_{(2,1)} = 0.25$; ICC$_{(2,k)} = 0.67$), suggesting that multi-seed aggregation can also yield stable estimates.
Given the relative reliability of aggregated ratings, we proceed to analyze the effects of prosodic manipulations on perceived sarcasm and naturalness for humans and models (see Figure~\ref{fig:human_machine_score}).

\subsection{Human naturalness perception}

We first examined the perceived naturalness associated with different prosodic manipulations. Results showed a significant main effect of speech rate ($\beta = 0.09$, $p < 0.001$), indicating that fast stimuli were perceived as more natural than slow stimuli. Loudness also showed a significant main effect ($\beta = 0.11$, $p < 0.001$), with soft stimuli receiving higher naturalness ratings than loud stimuli. Pitch contour did not show a significant main effect ($\beta = -0.04$, $p = 0.29$).
Importantly, significant interactions were observed between speech rate and pitch contour ($p < 0.05$) as well as between speech rate and loudness ($p < 0.05$), suggesting that the effect of temporal and intensity cues on perceived naturalness depends on their combination. In contrast, neither the pitch-contour-by-loudness interaction nor the three-way interaction reached statistical significance (both $p > 0.05$).
Although prosodic exaggeration reduced naturalness, overall ratings remained high for all conditions, suggesting that the stimuli were acceptable for sarcasm perception.

\subsection{Human sarcasm perception}

We then examined the perceived sarcasm associated with different prosodic manipulations. 
Results showed a significant main effect of loudness ($\beta = 0.29$, $p < 0.05$), indicating that loud stimuli received higher sarcasm ratings than soft stimuli. However, neither speech rate ($\beta = 0.06$, $p = 0.62$) nor pitch contour ($\beta = 0.14$, $p = 0.25$) showed significant main effects.
None of the two-way or three-way interactions reached statistical significance, indicating that the effects of speed, pitch contour, and loudness were additive and independent in the current setting. 

To further examine differences among the eight conditions, Tukey-adjusted pairwise comparisons were conducted based on estimated marginal means from the full factorial model. 
Comparisons indicated that conditions combining flat pitch contour and loud intensity were generally rated as more sarcastic than their soft counterparts. Specifically, the contrast between \textit{fast flat soft} and \textit{fast flat loud} was significant ($\beta = -0.24$, $p < 0.05$).
Similarly, \textit{slow flat soft} received significantly lower ratings than \textit{slow flat loud} ($\beta = -0.41$, $p < 0.05$). 
Across rate conditions, \textit{fast dynamic soft} was rated significantly lower than \textit{fast flat loud} ($\beta = -0.38$, $p < 0.05$) 
and \textit{slow flat loud} ($\beta = -0.49$, $p < 0.01$).  
Additional significant contrasts included differences between \textit{slow dynamic soft} and \textit{slow flat loud} ($\beta = -0.43$, $p < 0.01$), as well as several comparisons involving loud versus soft realizations within the same pitch configuration.
All remaining pairwise comparisons did not reach statistical significance after Tukey correction ($p > 0.05$).
Overall, these results suggest that increased loudness was the most reliable prosodic cue to perceived sarcasm, particularly when combined with flat pitch contours.
Optional participant comments were broadly consistent with this pattern: 
participants described \textit{fast dynamic soft} stimuli as sounding sincere and urgent, and characterized \textit{slow flat loud} stimuli as sounding sarcastic and angry.
This pattern aligns with previous research showing that prosodic exaggeration plays a central role in sarcasm perception \cite{rockwell2000lower, cheang_sound_2008}. 
Although pitch contour did not reach significance (p = 0.25), the direction of the effect is consistent with prior findings that intonational flattening contributes to perceived irony \cite{bryant2005there}. 
Taken together with the naturalness results, this pattern suggests that prosodic cues enhancing sarcasm perception (e.g., increased loudness) also introduce markedness that listeners interpret as reduced naturalness. 


\subsection{Machine naturalness perception}

For model-based speech perception, we first examined whether Qwen3-Omni perceived the synthesized prosodic manipulations differently in terms of naturalness.  
Results showed a significant main effect of pitch contour ($\beta = 0.12$, $p < 0.001$), indicating that dynamically modulated utterances were rated as more natural than flat-pitch utterances. 
In contrast, speech rate did not yield a significant main effect ($\beta = 0.08$, $p = 0.14$), nor did loudness ($\beta = 0.03$, $p = 0.54$). 
No significant two-way or three-way interactions were observed (all $p > 0.05$), suggesting that the contribution of pitch modulation to perceived naturalness was robust across different combinations of speech rate and intensity cues.
Although minor numerical differences were observed across prosodic conditions, overall naturalness ratings remained high in all cases (means ranging from 3.95 to 4.25), indicating that the acoustic 
manipulations preserved general perceptual acceptability.

\subsection{Machine sarcasm perception}
We further examined whether the synthesized prosodic manipulations influenced Qwen3-Omni's sarcasm ratings.
Results showed a significant main effect of speech rate on sarcasm ratings ($\beta = 0.31$, $p < 0.01$), indicating that slow stimuli were rated as more sarcastic than fast stimuli.
No significant main effects were observed for pitch contour ($\beta = 0.13$, $p = 0.27$) or loudness ($\beta = 0.04$, $p = 0.77$).
None of the two-way or three-way interactions reached statistical significance. 
Post-hoc pairwise comparisons with Tukey correction were conducted to examine differences across the eight prosodic conditions.
Significant contrasts were primarily observed between extreme prosodic configurations.
Conditions characterized by slow rate and loud intensity were consistently rated higher than fast-rate or soft-intensity conditions (e.g., $p < 0.01$ for \textit{slow dynamic loud} vs. \textit{fast dynamic soft}).
In contrast, comparisons between conditions that differed in only one prosodic dimension were generally non-significant after correction.
Overall, the pairwise results suggest that machine predictions are mainly driven by combinations of prosodic extremes rather than isolated features.

\subsection{Human-model alignment in sarcasm cue weighting}

To directly compare human and model sensitivity to prosodic manipulations, we examined the alignment between human sarcasm ratings and model-predicted sarcasm scores. A Spearman rank correlation analysis showed no significant rank-order alignment between human and model ratings across prosodic conditions ($\rho = -0.11$, $p = 0.26$).
Beyond this overall misalignment, systematic differences also emerged in how prosodic cues were weighted by the two systems (Table~\ref{tab:cue_weight_alignment}).

\begin{table}[h]
\centering
\caption{Comparison of prosodic cue weights between humans and the model.
Cue importance is measured by the fixed-effect coefficient ($\beta$) from the mixed-effects models.}
\begin{tabular}{lcc}
\toprule
Cue & Human $\beta$ & Machine $\beta$ \\
\midrule
Speech Rate (slow) & 0.061 & 0.313 \\
Pitch Variation (flat)       & 0.138 & 0.132 \\
Loudness (loud)     & 0.285 & 0.035 \\
\bottomrule
\end{tabular}
\label{tab:cue_weight_alignment}
\end{table} 

Rather than mirroring human sensitivity patterns, the model appeared to rely strongly on speech rate, whereas human listeners were primarily driven by loudness cues.
One possible explanation for such divergence is that human listeners acquire sarcasm sensitivity through socio-pragmatic experience, in which intensity cues are often associated with affective exaggeration. In contrast, the model’s behavior is shaped by large-scale multimodal training, where temporal regularities in speech may serve as more stable statistical signals. 
Pitch variation contributed less prominently in both systems, suggesting that this dimension plays a secondary role in the absence of contextual information. 

\section{Limitations and Future Work}
Our results clarify which prosodic dimensions primarily drive sarcasm perception under controlled synthetic manipulation. They also show limited behavioral alignment between human listeners and the model, which may reflect differences in the learning environments of humans and artificial systems.
However, these results should be interpreted with several limitations in mind. Because the stimuli were presented without rich discourse context, the task measures how acoustic manipulations bias perceived sarcasm rather than full comprehension of sarcastic intent. This context-free design may also explain why many conditions received mid-range ratings: listeners may have judged acoustic markedness rather than a clear pragmatic intention. In addition, many listeners were non-native English speakers, which may have affected cue weighting. Sarcasm was also treated as a single category, although irony and sarcasm are not identical and different sarcasm subtypes may rely on different cue combinations. Finally, while orthogonal manipulation isolates individual acoustic dimensions, natural prosodic cues often co-vary; separating them may therefore produce less natural speech. Future work could use context-rich stimuli, native-speaker populations, and more fine-grained sarcasm categories to improve ecological validity and generalizability.


\section{Generative AI Use Disclosure}
Generative AI tools were used for language editing and to enhance the clarity and readability of the manuscript. 
These tools were not used in the development of the research questions, theoretical framework, experimental design, data analysis, or interpretation of results, nor were they used to generate any substantive scientific content.

\bibliographystyle{IEEEtran}
\bibliography{mybib}

\end{document}